**Title**

Triple-sinusoid hedgehog lattice in a centrosymmetric Kondo metal

**Authors**

Soohyeon Shin,[1,2]† Jin-Hong Park,[3,2]† Romain Sibille,[4] Harim Jang,[2] Tae Beom Park,[2] Suyoung Kim,[5,2] Tian Shang,[6,1] Marisa Medarde,[1] Eric D. Bauer,[7] Oksana Zaharko,[4] Michel Kenzelmann,[4]* Tuson Park[2]*

**Affiliations**

[1]Laboratory for Multiscale Materials and Experiments, Paul Scherrer Institut, 5232 Villigen PSI, Switzerland.

[2]Center for Quantum Materials and Superconductivity (CQMS) and Department of Physics, Sungkyunkwan University, Suwon 16419, South Korea.

[3]Research Center for Novel Epitaxial Quantum Architectures, Department of Physics, Seoul National University, Seoul, 08826, South Korea.

[4]Laboratory for Neutron Scattering and Imaging, Paul Scherrer Institut, 5232 Villigen PSI, Switzerland.

[5]Department of Physics, Simon Fraser University, Burnaby, British Columbia, Canada.

[6]Key Laboratory of Polar Materials and Devices (MOE), School of Physics and Electronic Science, East China Normal University, Shanghai 200241, China.

[7]Los Alamos National Laboratory, Los Alamos, NM 87545, USA.

*Corresponding authors. E-mails: michel.kenzelmann@psi.ch (Michel Kenzelmann), tp8701@skku.edu (Tuson Park).

†These authors contributed equally to this work

**Abstract**

Superposed symmetry-equivalent magnetic ordering wave vectors can lead to topologically non-trivial spin textures, such as magnetic skyrmions and hedgehogs, and give rise to novel quantum phenomena due to fictitious magnetic fields associated with a non-zero Berry curvature of these spin textures. To date, all known spin textures are constructed through the superposition of multiple *spiral* orders, where spins vary in directions with constant amplitude. Recent theoretical studies have suggested that multiple *sinusoidal* orders, where collinear spins vary in amplitude, can construct distinct topological spin textures regarding chirality properties. However, such textures have yet to be experimentally realised. In this work, we report the observation of a zero-field magnetic hedgehog lattice from a superposition of triple sinusoidal wave vectors in the magnetically frustrated Kondo lattice $CePtAl_4Ge_2$. Notably, we also observe the emergence of anomalous electrical and thermodynamic behaviours near the field-induced transition from the zero-field topological hedgehog lattice to a non-topological sinusoidal state. These observations highlight the role of Kondo coupling in stabilising the zero-field hedgehog state in the Kondo lattice and warrant an expedited search for other topological magnetic structures coupled with Kondo coupling.

**Introduction**

Studies of *f*-electron heavy-fermion compounds have attracted significant interest because of rich quantum states, such as topological Kondo-insulators [1], unconventional superconductivity [2-4], quantum criticality [5,6], nematic phases [7-9], and topological magnetism [10-12]. With increasing hybridisation strength $J_{cf}$ between *f*- and conduction electrons, the nature of *f*-electrons changes from a localised to an itinerant character across the critical value of $J_c$, where non-Fermi-liquid (NFL) behaviour can be induced associated with magnetic quantum fluctuations [6,13]. Magnetically frustrated *f*-electron moments in metal are conducive to hosting a magnetically long-range ordered state by forming partially screened moments via the Kondo effect [14-16]. For instance, only 2/3 of localised Ce moments in CePdAl order antiferromagnetically, while 1/3 of them remain disordered. This mixed magnetic structure has been ascribed to competition between RKKY (Ruderman-Kittel-Kasuya-Yosida) interactions and Kondo effects in the magnetically frustrated lattice

[17-20]. Furthermore, applying a magnetic field along the magnetic-easy-axis suppresses partial Kondo screening of disordered Ce sites, giving rise to a possible quantum critical or spin liquid phase due to magnetic frustrations [18-22].

Magnetic skyrmions are defined by a non-zero skyrmion number $N_{sk}$ that counts how many times the local magnetisation direction wraps the unit sphere within the unit area. A magnetic hedgehog, a 3D topological spin texture, is defined by a hedgehog number $N_h = N_{sk}(-m_0) - N_{sk}(+m_0)$, where the $m_0$ is a uniform magnetisation [23]. Magnetic hedgehogs are constructed by radial swirling spins around a spin singular point, while skyrmions are vortex-like spin textures without a singular point. The primary mechanism for stabilising these swirling spin structures is commonly attributed to antisymmetric exchange interactions arising from the broken spatial inversion symmetry in noncentrosymmetric structures [24,25]. Nevertheless, two- or three-dimensional (2D or 3D) topological spin textures, magnetic skyrmions and hedgehogs, were discovered in centrosymmetric $f$-electron magnets, raising the possibility that they may arise from the competition between frustrated RKKY interactions and easy-axis spin anisotropy [10,11,26,27]. Furthermore, it was proposed that magnetic skyrmions arise from the 2D frustrated RKKY interactions, while the 3D frustrated RKKY interactions lead to magnetic hedgehogs [25].

In a reciprocal space, skyrmion and hedgehog lattices are depicted as superpositions of multiple magnetic orderings in 2D and 3D reciprocal spaces, respectively. So far, all known topological spin textures are constructed through a superposition of multiple spiral orderings where spins vary in directions with constant amplitude. Recently, effective spin models derived from the Kondo lattice model have shown that the topological spin textures can also be stabilised by a superposition of multiple sinusoidal orderings where collinear spins vary in amplitude and exhibit different spin chirality properties [28]. Moreover, these effective spin models have suggested the emergence of a zero-field hedgehog lattice (HL) and the importance of the higher-order RKKY interactions enhanced by nesting bands [29,30]. In this work, we report evidence for a zero-field magnetic hedgehog lattice in the Kondo metal $CePtAl_4Ge_2$, the first topological spin texture composed of multiple sinusoidal orderings to the best of the author's knowledge. In addition, we suggest this hedgehog lattice as a model system for examining the effective spin model.

**Results**

The *f*-electron frustrated antiferromagnet CePtAl$_4$Ge$_2$, in which localised Ce moments form a quasi-2D triangular lattice, exhibits an antiferromagnetic (AFM) state stabilised through RKKY interactions below $T = 2.3$ K ($T_N$) [31]. Neutron scattering in zero magnetic field experiments revealed an incommensurate order with a wave vector of **k** = (1.39, 0, 0.09) described by the $\Gamma_2$ irreducible representation (IR), where the moments are parallel to **k** [32]. The single-**k** structure corresponds to a sinusoidal AFM (1**k**-sin), where about 30 % of ordered moments remain disordered, as shown in Fig. 1a. The 1**k**-sin state may reflect the intricate competition between RKKY interactions and the Kondo effect in the magnetically frustrated triangular lattice. Our spin simulation reveals that the multi-**k** model in which the symmetry-equivalent three arms of **k** = (1.39, 0, 0.09) are superposed corresponds to a triple-**k** hedgehog lattice (3**k**-HL), as shown in Fig. 1b (for details, see Figs. S1-S4 in section I of supplementary information, SI), indicating that CePtAl$_4$Ge$_2$ could be a candidate topological magnet with competition between frustrated RKKY and Kondo interactions.

Figure 1c shows that field-dependent electrical resistivity $\rho(H)$ of CePtAl$_4$Ge$_2$ at $T = 0.6$ K exhibits two successive sharp changes at $H = 6.7$ and 9.2 kOe designated by $H_L^*$ and $H_U^*$, respectively, when the field is applied along the crystallographic [010] direction ($H_{[010]}$). Within the experimental resolution, $\rho(H)$ at $T = 0.6$ K are identical for increasing and decreasing fields, suggesting that these are second or weak first-order phase transitions. These critical fields, $H_L^*$ and $H_U^*$, decrease with increasing temperature (see Fig. S5 in SI). Field-dependent magnetisation $M(H)$ at $T = 0.6$ K (see Fig. S6 in SI) shows weak jumps at these two critical fields, which are manifested as distinct peaks in $\partial M/\partial H$. This demonstrates that these transitions are of magnetic origin. We will show below that the sharp resistivity changes at critical fields correspond to a magnetic phase transition from 3**k**-HL to 1**k**-sin with increasing field. As observed in CePtAl$_4$Ge$_2$, topological magnetic phases, such as hedgehogs in MnGe [33] and skyrmions in MnSi [34] and Gd$_2$PdSi$_3$ [35], exhibit higher electrical resistivity than adjoining non-topological magnetic phases, attributed to the additional spin scattering from fictitious magnetic fields of topological spin textures [23].

The drop of $\rho(H)$, shown in Fig. 1d as negative peaks in the first derivative of electrical resistivity $\partial\rho/\partial H$, can be ascribed to the suppression of the topological defects.

Single-crystal neutron diffraction experiments were performed to investigate the predicted zero-field HG phase by the spin simulation and the field-induced magnetic phase transition. As shown in Fig. 2a, three magnetic reflections were observed at zero-field corresponding to symmetry-equivalent three arms of **k** in the hexagonal base of $R\bar{3}m$, i.e., $\mathbf{k}_1 = (1.39, 0, 0.09)$, $\mathbf{k}_2 = (0, -1.39, 0.09)$, and $\mathbf{k}_3 = (-1.39, 1.39, 0.09)$. Our structural refinement shows that the multi-arm magnetic structure may be interpreted as either a multi-**k** single-domain or a single-**k** multi-domain structure [36]. Indeed, the zero-field magnetic reflections result in a similar fit quality of refinement using the multi-**k** and the multi-domain structure (see Table S1 in SI). Figure 2b shows the symmetry-equivalent three **k**-arms projected on the ($hk0$)-plane and three sinusoidal ordering configurations. When we applied a magnetic field of $H_{[010]} = 15$ kOe, only $\mathbf{Q}(\mathbf{k}_1)$ is observed because the modulating direction of the $\mathbf{k}_1$-arm is perpendicular to $H_{[010]}$, as shown in Fig. 2a, (for details, see Fig. S7 in SI).

Figure 2c shows that $H_{[010]}$-dependent neutron intensities of the three **k**-arms, i.e., $\mathbf{Q}(\mathbf{k}_1) = (0.61, 0, -4.09)$, $\mathbf{Q}(\mathbf{k}_2) = (1, -0.39, 3.09)$, and $\mathbf{Q}(\mathbf{k}_3) = (0.61, 0.39, 3.09)$, at $T = 0.15$ K, where the intensities are same in the low-field regime. As $H_{[010]}$ increases, the neutron intensity of the three arms differ above $H = 5.5(5)$ kOe; the neutron intensity of $\mathbf{Q}(\mathbf{k}_1)$ increases while the others decrease before completely disappearing around $H = 10$ kOe. These two fields are consistent with $H_L^*$ and $H_U^*$, respectively. At $H \geq H_U^*$, only magnetic reflections of the $\mathbf{k}_1$-arm were observed and were well refined by a $\Gamma_2$ IR (details in Fig. S7 and Table S1 in SI).

We show that the field dependence of the Bragg peak intensity of the three arms provides evidence that the magnetic structure is multi-**k** at zero-field and does not form **k**-domains. In Fig. 2c, field-dependent neutron intensity at $\mathbf{Q}(\mathbf{k}_3)$ was measured with decreasing field, while $\mathbf{Q}(\mathbf{k}_1)$ and $\mathbf{Q}(\mathbf{k}_2)$ were measured with increasing field. The emergence of neutron intensity at $\mathbf{Q}(\mathbf{k}_3)$ with decreasing field indicates a correlation among the three **k**-arms. If the low-field phase consists of different single-**k** domains, the magnetic Bragg peak at

$Q(k_3)$ would not increase in intensity at $H \leq H_U^*$ as a function of decreasing $H_{[010]}$. The zero-field neutron intensities of three arms are recovered at $H \leq H_L^*$ regardless of field-sweep direction, the so-called lock-in transition. The field-dependent neutron intensities of the three **k**-arms are in good agreement with not only the model calculation of the magnetic hedgehog system established by the triple-**k** structure [29] but also the single-crystal neutron results of the hedgehog lattice SrFeO$_3$ [27]. Supporting the conclusion, the field-dependent neutron intensities do not show a reasonable quantitative agreement between the increase in $Q(k_1)$ and the decreases in others, contrary to what would be expected from the multi-domain structures [37-40]. Note that **k** remains unchanged across the field-induced transitions (see Fig. S8 in SI).

The triple-**k** Γ$_2$ IR ($H = 0$) and single-**k** Γ$_2$ IR ($H_{[010]} = 15$ kOe) magnetic structures obtained from the refinements are illustrated in Figs. 2d and 2e, respectively, wherein antiferromagnetically ordered moments (red arrows) are non-coplanar in the triple-**k** state but are collinear in the single-**k** state. The experimentally obtained triple-**k** structure, in which three sinusoidal ordering are superposed, is consistent with the simulated HL (for details, see Fig. S9 in SI), underlining that the zero-field HL is stabilised in CePtAl$_4$Ge$_2$. The best refinements for $H = 0$ and $H_{[010]} = 15$ kOe are shown in Fig. S7 and Table S1 in SI. The magnetic structure of the intermediate phase (I), in which three arms are unequally superposed, might be a distorted HL whose symmetry is different from the zero-field HL. The specific heat capacity divided by temperature $C/T$ exhibits a peak at the HL-to-I phase boundary as an indication of the symmetry change (see Fig. S9a).

Figure 3 illustrates the temperature-dependent electrical resistivity $\rho(T)$ of CePtAl$_4$Ge$_2$ at various $H_{[010]}$. Figure 3a shows that $\rho(T)$, plotted in the square of temperature $T^2$, exhibits $T^2$ behaviour at temperatures below a crossover temperature $T_{e-e}$, below which the electron-electron scattering is dominant. Solid red lines show the best least-squares fitting results using $\rho(T) = \rho_0 + AT^2$ at $T \leq T_{e-e}$. As $H_{[010]}$ increases, $T_{e-e}$ decreases, showing a minimum at around $H = 10$ kOe, before increasing with further increasing the magnetic field. As shown in Fig. 3b, the minimum of $T_{e-e}$ (dashed violet line) can be attributed to the additional scattering source near $H_U^*$, where the magnetic phase transition occurs from the 3**k**-HL to the 1**k**-sin. Figure 3c shows that the field dependence of the temperature-square coefficient

$A(H)$, deduced from the least-squares fittings, sharply increases from 0.369 to 0.503 $\mu\Omega\cdot$cm/K$^2$ near $H=10$ kOe. The anomaly near $H_U^*$ is also shown in the bulk property, as shown in Fig. 3d. Magnetic entropy $S_{4f}$ at $T = 1$ K increases with increasing field due to field-induced liberated magnetic moments through the suppression of the Kondo screening [19]. Note that $S_{4f} = \int C_{4f}/T\, dT$, when $C_{4f} = C(\text{CePtAl}_4\text{Ge}_2) - C(\text{LaPtAl}_4\text{Ge}_2)$. About 30 % screened moments at zero-field are fully liberated between $H_L^*$ and $H_U^*$ where $M(H)$ is 0.35 $\mu_B$ (~1/3 of the saturated magnetisation; see also Fig. S6 in SI). With this observation of the enhanced quasiparticle scattering across the 3**k**-to-1**k** phase transition, we suggest abundant magnetic fluctuations emerging from the frustrated metallic ground state. The fact that $A(H)$ changes sharply near $H_U^*$, where $S_{4f}$ is at the maximum, indicates that the source of the additional scattering is related to the field-induced magnetic fluctuation driven by non-ordered moments in the AFM of CePtAl$_4$Ge$_2$.

Figure 4a shows an overall $T - H$ magnetic phase diagram of CePtAl$_4$Ge$_2$ overlaid on a colour map of the temperature exponent $n$, $n = \partial \ln(\rho - \rho_0)/\partial \ln T$. In the phase diagram, a magenta dashed line, representing the linear extrapolation of Zeeman characteristic temperatures, penetrates the AFM phase boundary $H_c$, and approaches the HL boundary, $H_L^*$. Figure 4b shows three representative $C_{4f}/T$ for different field regimes. While zero-field $C_{4f}/T$ exhibits a peak due to the magnetic phase transition, $C_{4f}/T$ shows an additional broad peak $T^*$ (half-filled squares) within the AFM phase at $H = 15$ kOe that corresponds to the Schottky anomaly at $H = 30$ kOe (for other fields see Fig. S9 in SI). The characteristic temperature of the Schottky-type broad peak (= $T_Z$) in $C_{4f}/T$ is proportional to $H$ as the crystal field $|\pm 1/2\rangle$ doublet is split by the Zeeman effect [32]. In the field-polarized ferromagnetic state (FFM, $H > H_c$), as shown in Fig. 4c, $T_Z$ decreases to $H = 30$ kOe with a decreasing field. With further lowering field, as observed at $H = 15$ kOe, the broad peak of $C_{4f}/T$ is observed even deep in the AFM state following the extrapolated $T_Z$ line (magenta dashed line). The red colour of $n$, $n\sim3$, also penetrates the AFM phase boundary following the extrapolated $T_Z$ line, indicating that the $T^3$-dependent resistivity can be attributed to the Zeeman effect of the non-ordered moments. The crossover field $H_Z(T = 0) \sim 3.5$ kOe is the minimum field required for a Schottky anomaly and, as shown in Fig. 4c, indicates a possible suppression of Kondo screening even in the AFM phase ($H < H_c$). These results underline that Zeeman effects on the non-ordered $4f^1$-doublet are important in the AFM phase because any field larger than $H_Z(0)$ suppresses Kondo screening. The proximity between $H_Z(0)$ and $H_L^*$ raises the possibility that the transition from the topological 3**k**-HL to the non-topological 1**k**-sin state is correlated with the Kondo coupling in the magnetically frustrated lattice.

**Discussion**

The significant role of Kondo coupling in stabilising topological spin textures has been suggested through effective spin Hamiltonians derived from the Kondo model [29,30,41-43]. These Hamiltonians highlight the influence of higher-order RKKY interactions, which favour noncoplanar spin textures and are sensitive to the filling factor of electron band [41,43]. For instance, a triangular Kondo lattice can exhibit noncoplanar spin textures with a finite spin chirality when the filling factor is around 1/4 and 3/4. Fermi surface nesting, driven by these specific band-filling factors, enhances $4^{th}$-order RKKY interactions, resulting in multi-**k** noncoplanar magnetic structures. Therefore, by performing angle-resolved photoemission x-ray spectroscopy experiments on CePtAl$_4$Ge$_2$, it is possible to confirm whether the HL is associated with the Fermi surface nesting. Furthermore, an investigation into the magnetic structure of the isostructural compound CeAuAl$_4$Ge$_2$ [44] can provide evidence of how the band-filling factor influences the HL in CePtAl$_4$Ge$_2$, given that Au contributes fewer or more electrons to the system.

Hedgehog lattices of cubic systems, exemplified by MnGe [25] and SrFeO$_3$ [27], consist of an alternating arrangement of monopoles and antimonopoles uniformly distributed in 3D space. However, in the case of CePtAl$_4$Ge$_2$, one would anticipate anisotropic monopole/antimonopole interactions due to the quasi-2D triangular Ce layers. Furthermore, the observation of the HL in CePtAl$_4$Ge$_2$ represents the first experimental realisation of a topological spin texture resulting from the superposition of multiple sinusoids. Up to this point, all reported topological magnetic textures are driven by multiple spirals [10-12,24,25,27,45]. As demonstrated in previous theoretical studies [28], the collective net spin chirality in multi-sinusoid noncoplanar spin textures is nullified since local spin chirality cancels each other out, resulting in the absence of topological Hall effects. To validate the existence of topological spin textures, the observation of the topological Hall effects is considered strong evidence since the spin textures, driven by the superposition of multiple ordering wave vectors, depend on the phase factors of these wave vectors [10,25,28]. Nevertheless, the presence of the HL in $CePtAl_4Ge_2$ remains valid even in the absence of topological Hall effects, as the triple-sinusoid HLs do not rely on phase shifts of the sinusoidal ordering wave vectors [28].

In summary, we have reported the stabilisation of a magnetic hedgehog lattice, driven by the superposition of triple-**k** sinusoidal ordering wave vectors, in the magnetically frustrated Kondo metal. Field-induced spin fluctuations and anomalous electronic transport behaviour are observed across the topological phase transition from the topological hedgehog lattice to the non-topological sinusoidal antiferromagnetic state. These discoveries warrant an expedited search for other topological magnetic structures coupled with Kondo coupling and for novel physical phenomena arising from anisotropic monopole/antimonopole interactions within quasi-2D triangular lattices.

**References**

1   Neupane, M. *et al.* Surface electronic structure of the topological Kondo-insulator candidate correlated electron system $SmB_6$. *Nature Communications* **4**, 2991 (2013). https://doi.org:10.1038/ncomms3991

2   Steglich, F. *et al.* Superconductivity in the Presence of Strong Pauli Paramagnetism: $CeCu_2Si_2$. *Physical Review Letters* **43**, 1892-1896 (1979). https://doi.org:10.1103/PhysRevLett.43.1892

3   Kenzelmann, M. *et al.* Coupled Superconducting and Magnetic Order in $CeCoIn_5$. *Science* **321**, 1652-1654 (2008). https://doi.org:10.1126/science.1161818

4   Jiao, L. *et al.* Chiral superconductivity in heavy-fermion metal $UTe_2$. *Nature* **579**, 523-527 (2020). https://doi.org:10.1038/s41586-020-2122-2

5   Stewart, G. R. Non-Fermi-liquid behavior in *d*- and *f*-electron metals. *Reviews of Modern Physics* **73**, 797-855 (2001). https://doi.org:10.1103/RevModPhys.73.797

6   Park, T. *et al.* Hidden magnetism and quantum criticality in the heavy fermion superconductor $CeRhIn_5$. *Nature* **440**, 65-68 (2006). https://doi.org:10.1038/nature04571

7   Park, T. *et al.* Textured Superconducting Phase in the Heavy Fermion $CeRhIn_5$. *Physical Review Letters* **108**, 077003 (2012). https://doi.org:10.1103/PhysRevLett.108.077003

8   Fobes, D. M. *et al.* Tunable emergent heterostructures in a prototypical correlated metal. *Nature Physics* **14**, 456-460 (2018). https://doi.org:10.1038/s41567-018-0060-9

9   Rosa, P. F. S. *et al.* Enhanced Hybridization Sets the Stage for Electronic Nematicity in $CeRhIn_5$. *Physical Review Letters* **122**, 016402 (2019). https://doi.org:10.1103/PhysRevLett.122.016402

10  Kurumaji, T. *et al.* Skyrmion lattice with a giant topological Hall effect in a frustrated triangular-lattice magnet. *Science* **365**, 914 (2019). https://doi.org:10.1126/science.aau0968

11  Yasui, Y. *et al.* Imaging the coupling between itinerant electrons and localised moments in the centrosymmetric skyrmion magnet $GdRu_2Si_2$. *Nat Commun* **11**, 5925 (2020). https://doi.org:10.1038/s41467-020-19751-4

12  Puphal, P. *et al.* Topological Magnetic Phase in the Candidate Weyl Semimetal CeAlGe. *Physical Review Letters* **124**, 017202 (2020). https://doi.org:10.1103/PhysRevLett.124.017202

13  Park, T. *et al.* Isotropic quantum scattering and unconventional superconductivity. *Nature* **456**, 366-368 (2008). https://doi.org:10.1038/nature07431

14  Lacroix, C. Frustrated Metallic Systems: A Review of Some Peculiar Behavior. *Journal of the Physical Society of Japan* **79** (2010). https://doi.org:Artn 011008 10.1143/Jpsj.79.011008

15  Sato, T., Assaad, F. F. & Grover, T. Quantum Monte Carlo Simulation of Frustrated Kondo Lattice Models. *Physical Review Letters* **120**, 107201 (2018). https://doi.org:10.1103/PhysRevLett.120.107201

16  Motome, Y., Nakamikawa, K., Yamaji, Y. & Udagawa, M. Partial Kondo Screening in Frustrated Kondo Lattice Systems. *Physical Review Letters* **105**, 036403 (2010). https://doi.org:10.1103/PhysRevLett.105.036403

17  Dönni, A. *et al.* Geometrically frustrated magnetic structures of the heavy-fermion compound CePdAl studied by powder neutron diffraction. *Journal of Physics: Condensed Matter* **8**, 11213-11229 (1996). https://doi.org:10.1088/0953-8984/8/50/043

18  Oyamada, A. *et al.* Critical behavior in a Kondo-screening partially-ordered antiferromagnet CePdAl. *Journal of Physics: Conference Series* **320**, 012067 (2011). https://doi.org:10.1088/1742-6596/320/1/012067

19  Lucas, S. *et al.* Entropy Evolution in the Magnetic Phases of Partially Frustrated CePdAl. *Physical Review Letters* **118**, 107204 (2017). https://doi.org:10.1103/PhysRevLett.118.107204

20  Fritsch, V. *et al.* CePdAl - a Kondo lattice with partial frustration. *Journal of Physics: Conference Series* **807**, 032003 (2017). https://doi.org:10.1088/1742-6596/807/3/032003

21  Zhang, J. *et al.* Kondo destruction in a quantum paramagnet with magnetic frustration. *Physical Review B* **97**, 235117 (2018). https://doi.org:10.1103/PhysRevB.97.235117

22  Zhao, H. C. *et al.* Quantum-critical phase from frustrated magnetism in a strongly correlated metal. *Nature Physics* **15**, 1261-1266 (2019). https://doi.org:10.1038/s41567-019-0666-6

23  Han, J. H. *Skyrmions in condensed matter*. Vol. 278 (Springer, 2017).

24  Mühlbauer, S. *et al.* Skyrmion Lattice in a Chiral Magnet. *Science* **323**, 915-919 (2009). https://doi.org:10.1126/science.1166767

25  Fujishiro, Y. *et al.* Topological transitions among skyrmion- and hedgehog-lattice states in cubic chiral magnets. *Nat Commun* **10**, 1059 (2019). https://doi.org:10.1038/s41467-019-08985-6

26  Hayami, S., Lin, S.-Z. & Batista, C. D. Bubble and skyrmion crystals in frustrated magnets with easy-axis anisotropy. *Physical Review B* **93**, 184413 (2016). https://doi.org:10.1103/PhysRevB.93.184413

27  Ishiwata, S. *et al.* Emergent topological spin structures in the centrosymmetric cubic perovskite $SrFeO_3$. *Physical Review B* **101**, 134406 (2020). https://doi.org:10.1103/PhysRevB.101.134406

28  Shimizu, K., Okumura, S., Kato, Y. & Motome, Y. Phase degree of freedom and topology in multiple-Q spin textures. *Physical Review B* **105** (2022). https://doi.org:10.1103/PhysRevB.105.224405

29  Okumura, S., Hayami, S., Kato, Y. & Motome, Y. Magnetic hedgehog lattices in noncentrosymmetric metals. *Physical Review B* **101**, 144416 (2020). https://doi.org:10.1103/PhysRevB.101.144416

30  Okumura, S., Hayami, S., Kato, Y. & Motome, Y. Magnetic Hedgehog Lattice in a Centrosymmetric Cubic Metal. *Journal of the Physical Society of Japan* **91** (2022). https://doi.org:10.7566/jpsj.91.093702

31  Shin, S. *et al.* Synthesis and characterization of the heavy-fermion compound $CePtAl_4Ge_2$. *Journal of Alloys and Compounds* **738**, 550-555 (2018). https://doi.org:10.1016/j.jallcom.2017.12.180

32  Shin, S. *et al.* Magnetic structure and crystalline electric field effects in the triangular antiferromagnet $CePtAl_4Ge_2$. *Physical Review B* **101**, 224421 (2020). https://doi.org:10.1103/PhysRevB.101.224421

33  Kanazawa, N. *et al.* Large Topological Hall Effect in a Short-Period Helimagnet MnGe. *Physical Review Letters* **106**, 156603 (2011). https://doi.org:10.1103/PhysRevLett.106.156603

34  Yokouchi, T. *Magneto-transport Properties of Skyrmions and Chiral Spin Structures in MnSi*.  (Springer Singapore, 2019).

35　Zhang, H. *et al.* Anomalous magnetoresistance in centrosymmetric skyrmion-lattice magnet $Gd_2PdSi_3$. *New Journal of Physics* **22**, 083056 (2020). https://doi.org:10.1088/1367-2630/aba650

36　Price, D. L. & Sköld, K. *Neutron scattering.. Part C*.  (Academic Press, 1987).

37　Herrmann-Ronzaud, D., Burlet, P. & Rossat-Mignod, J. Equivalent type-II magnetic structures: CoO, a collinear antiferromagnet. *Journal of Physics C: Solid State Physics* **11**, 2123-2137 (1978). https://doi.org:10.1088/0022-3719/11/10/023

38　White, J. S. *et al.* Stress-induced magnetic domain selection reveals a conical ground state for the multiferroic phase of $Mn_2GeO_4$. *Physical Review B* **94**, 024439 (2016). https://doi.org:10.1103/PhysRevB.94.024439

39　Lhotel, E. *et al.* Field-induced phase diagram of the XY pyrochlore antiferromagnet $Er_2Ti_2O_7$. *Physical Review B* **95**, 134426 (2017). https://doi.org:10.1103/PhysRevB.95.134426

40　Fabrèges, X. *et al.* Field-driven magnetostructural transitions in $GeCo_2O_4$. *Physical Review B* **95**, 014428 (2017). https://doi.org:10.1103/PhysRevB.95.014428

41　Akagi, Y. & Motome, Y. Spin Chirality Ordering and Anomalous Hall Effect in the Ferromagnetic Kondo Lattice Model on a Triangular Lattice. *Journal of the Physical Society of Japan* **79**, 083711 (2010). https://doi.org:10.1143/JPSJ.79.083711

42　Hayami, S., Ozawa, R. & Motome, Y. Effective bilinear-biquadratic model for noncoplanar ordering in itinerant magnets. *Physical Review B* **95**, 224424 (2017). https://doi.org:10.1103/PhysRevB.95.224424

43　Martin, I. & Batista, C. D. Itinerant Electron-Driven Chiral Magnetic Ordering and Spontaneous Quantum Hall Effect in Triangular Lattice Models. *Physical Review Letters* **101**, 156402 (2008). https://doi.org:10.1103/PhysRevLett.101.156402

44　Zhang, S. *et al.* Electronic structure and magnetism in the layered triangular lattice compound $CeAuAl_4Ge_2$. *Physical Review Materials* **1**, 044404 (2017). https://doi.org:10.1103/PhysRevMaterials.1.044404

45　Nomoto, T., Koretsune, T. & Arita, R. Formation Mechanism of the Helical Q Structure in Gd-Based Skyrmion Materials. *Phys Rev Lett* **125**, 117204 (2020). https://doi.org:10.1103/PhysRevLett.125.117204

46　　Rodríguez-Carvajal, J. Recent advances in magnetic structure determination by neutron powder diffraction. *Physica B: Condensed Matter* **192**, 55-69 (1993). https://doi.org:https://doi.org/10.1016/0921-4526(93)90108-I

47　　Schotte, K. D. & Schotte, U. Interpretation of Kondo experiments in a magnetic field. *Physics Letters A* **55**, 38-40 (1975). https://doi.org:https://doi.org/10.1016/0375-9601(75)90386-2

**Materials and Methods**

Single-crystal $CePtAl_4Ge_2$ was grown by a self-flux method, whose details are described elsewhere [31]. Electrical resistivity was measured using the conventional four-probe technique by applying a current of 1 mA and a magnetic field along the crystallographic [010] direction. Magnetization was measured using a superconducting quantum interference device in a Quantum Design Magnetic Property Measurement System (MPMS-XL, Quantum Design) with the He-3 option. The heat capacity was measured using the two-tau thermal relaxation method. Neutron diffraction experiments were performed on a single piece of single crystal using a thermal neutron diffractometer Zebra at SINQ. Structural refinements of the neutron reflections were performed using the FullProf suite [46]. The single crystal was aligned horizontally in the $(h,0,l)$ reciprocal plane on Zebra, and neutrons with $\lambda$ = 2.32 and 1.17 Å were adopted. The sample holder was attached to the mixing chamber of an Oxford dilution refrigerator for monitoring the magnetic diffraction down to 0.1 K. The magnetic ground states of $CePtAl_4Ge_2$ were calculated in the context of the multi-**k** spin model to ensure the stability of the spin structure. Through the multi-**k** spin model, the magnetic structures were visualized.

**Acknowledgments**

The authors would like to thank Dr. Vladimir Pomjakushin, Dr. Juan Rodriguez-Carvajal, and Dr. Jonathan S. White for the fruitful discussion about the magnetic structure.

**Funding:**

National Research Foundation of Korea grant funded by the Korean Ministry of Science, ICT, and Planning No. 2021R1A2C2010925 (TP)

National Research Foundation of Korea grant funded by the Korean Ministry of Education No. NRF-2019R1A6A1A10073079 (TP)

National Research Foundation of Korea grant funded by the Korean Ministry of Education No. NRF-2020R1I1A1A01073543 (JP)

"Young Researchers Exchange Programme between Korea and Switzerland" under the "Korean-Swiss Science and Technology Programme 2017-2018" (SS, MK)

Swiss National Science Foundation Project No. 200021_88706 (SS)

**Author contributions:**

    Conceptualisation: TP, MK

    Synthesis: SS, EDB

    Electrical transport: SS, SK, TBP

    Heat capacity: SS, HJ

    Magnetisation: SS, TS, MM

    Neutron scattering: SS, RS, OZ

    Writing—original draft: SS, JP, TP, MK

    Writing—review & editing: RS, MM, HJ, EDB

**Competing interests:**

Authors declare that they have no competing interests.

**Data and materials availability:**

The data supporting this study are available via the Zenodo repository (DOI:10.5281/zenodo.6418591).

**Figures and captions**

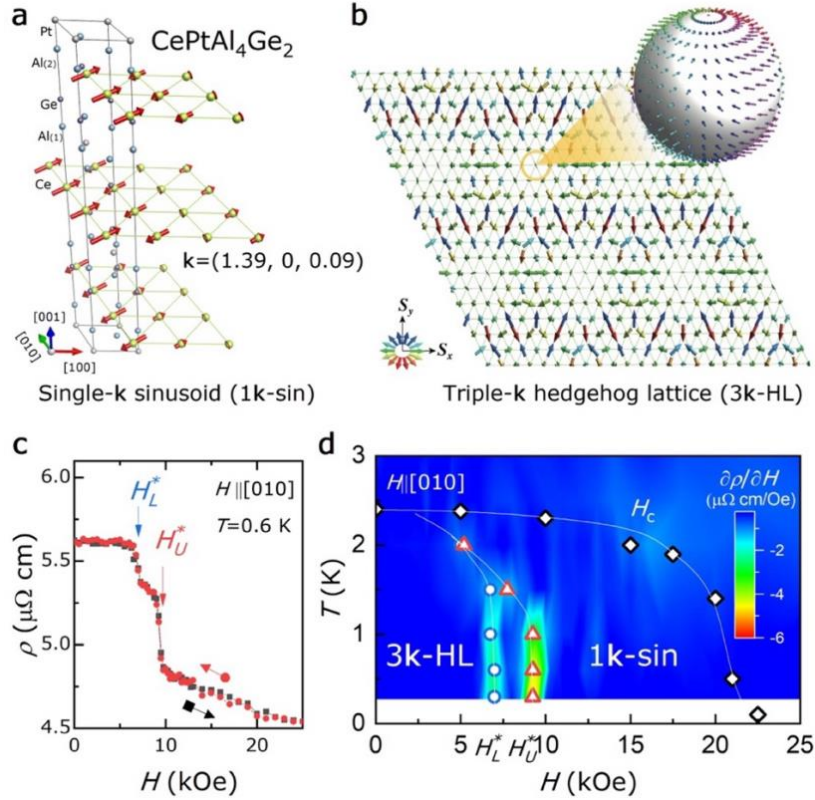

**Fig. 1. Magnetic structures and temperature−magnetic field ($T-H$) phase diagram of CePtAl$_4$Ge$_2$. a**, In CePtAl$_4$Ge$_2$, ordered magnetic moments of Ce$^{3+}$ modulate with a propagation vector **k** = (1.39, 0, 0.09), so-called sinusoidal ordering. The black solid line indicates the crystal structure unit cell ($R\bar{3}m$, no. 166) in a hexagonal base. Red arrows and green spheres represent ordered magnetic moments and Ce atoms. Blue, grey, and violet spheres represent Al, Pt, and Ge atoms. **b**, The hedgehog lattice (HL) spin texture is obtained by superpositioning the three symmetry-equivalent sinusoidal orderings. **c**, Field-dependent electrical resistivity $\rho(H)$ was measured at $T$ = 0.6 K with increasing (close black squares) and decreasing (open red circles) magnetic fields whose direction is along the crystallographic [010] direction. The low- and high-field resistivity drops with increasing field are indicated by $H_L^*$ and $H_U^*$, respectively. **d**, A colour contour plot of the first derivative of electrical resistivity $\partial\rho/\partial H$ shows the field-induced magnetic phase transition from the topological 3**k**-HL to the non-topological 1**k**-sin state across the critical fields of $H_L^*$ and $H_U^*$. The red colour means the abrupt suppression of electrical resistivity with increasing $H$, indicating the existence of strong electronic scattering sources in 3**k**-HL. $H_c$ indicates the phase boundary between the magnetically ordered and the paramagnetic states.

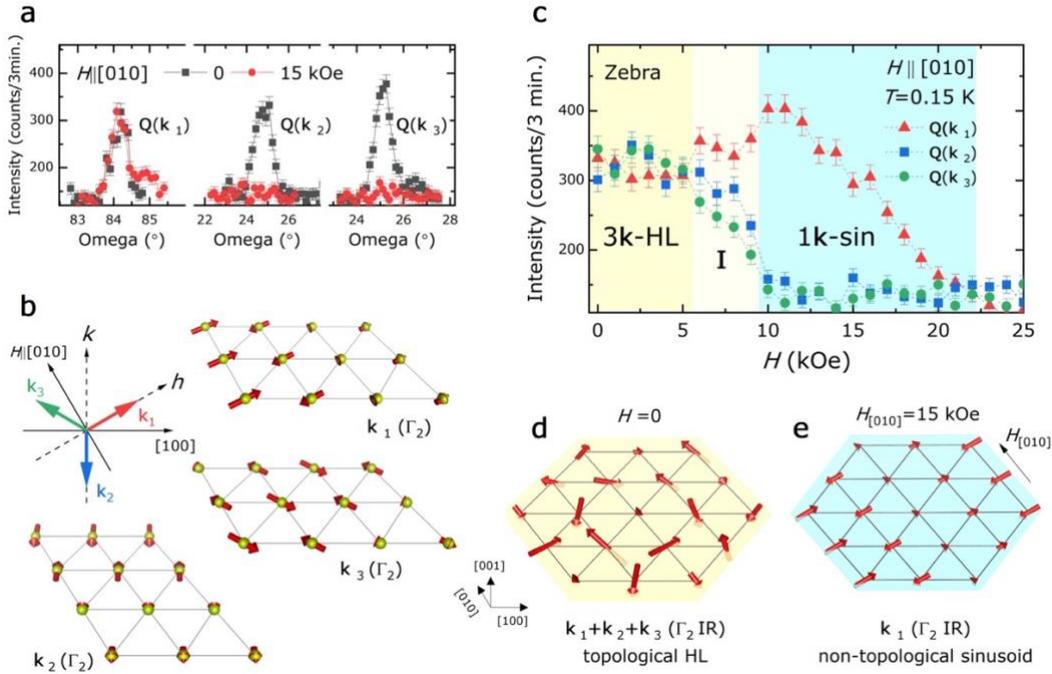

**Fig. 2. Single-crystal neutron diffraction results of CePtAl$_4$Ge$_2$. a**, Omega scan of three magnetic reflections, *i.e.*, **Q**(**k**$_1$) = (0.61, 0, -4.09) = (2, 0, -4) − **k**$_1$, **Q**(**k**$_2$) = (1, -0.39, 3.09) = (1, 1, 3) + **k**$_2$, and **Q**(**k**$_3$) = (0.61, 0.39, 3.09) = (2, -1, 3) + **k**$_3$, at zero-field (Black squares) and $H_{[010]}$ = 15 kOe (Red circles). **b**, Symmetry-equivalent three arms of **k** are projected on the (*hk*0)-plane. The solid (dashed) black arrows show the crystallographic (reciprocal) axes. *H*∥[010] indicates the direction of the applied magnetic field perpendicular to the **k**$_1$-arm. The three magnetic structures show symmetry-equivalent sinusoidal orderings with $\Gamma_2$ irreducible representation (IR). **c**, Neutron intensity of the three magnetic reflections representing different **k**-arms was monitored at *T* = 0.15 K as a function of $H_{[010]}$. The yellow (*H* < 5.5 kOe), transparent yellow (5.5 ≤ H < 9.5 kOe), and cyan (*H* ≥ 9.5 kOe) colour regions represent the triple-**k** hedgehog lattice (3**k**-HL) with $\Gamma_2$ IR, intermediate states (I), and the single-**k** sinusoidal ordering (1**k**-sin) with $\Gamma_2$ IR, respectively. Magnetic structures obtained from the refinement of single-crystal neutron diffraction results at the zero-field and $H_{[010]}$ = 15 kOe are described in **d** and **e,** respectively. Red arrows represent the ordered magnetic moments.

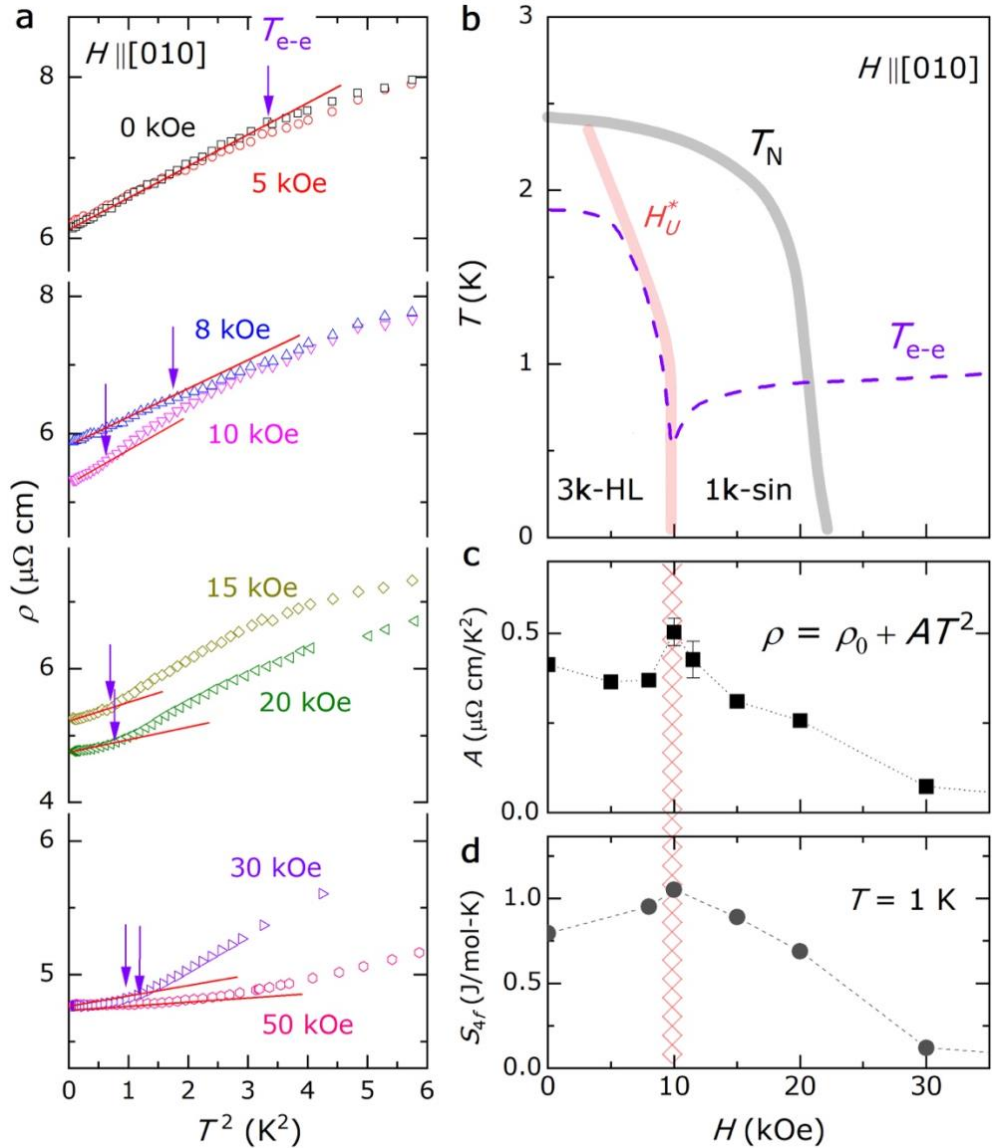

**Fig. 3. Enhanced electronic scattering near the field-induced magnetic phase transition of CePtAl$_4$Ge$_2$. a**, $T^2$-dependent electrical resistivity results under various magnetic fields, applied along the [010] direction ($H_{[010]}$), are displayed in four panels. Each solid red line is the least-squares fitting using $\rho(T) = \rho_0 + AT^2$ below temperature $T_{e\text{-}e}$ where the scattering mechanism is mainly governed by electron-electron scattering. **b**, The estimated $T_{e\text{-}e}$ (dashed violet line) overlaid on the simplified $T - H$ phase diagram. **c**, The $T^2$-coefficient $A$ deduced from the least-squares fitting is depicted as a function of $H_{[010]}$. **d**, Magnetic entropy ($S_{4f}$) at $T = 1$ K is plotted as a function of $H_{[010]}$. The hashed bar in **c** and **d** indicates the maximum of $A$ and $S_{4f}$.

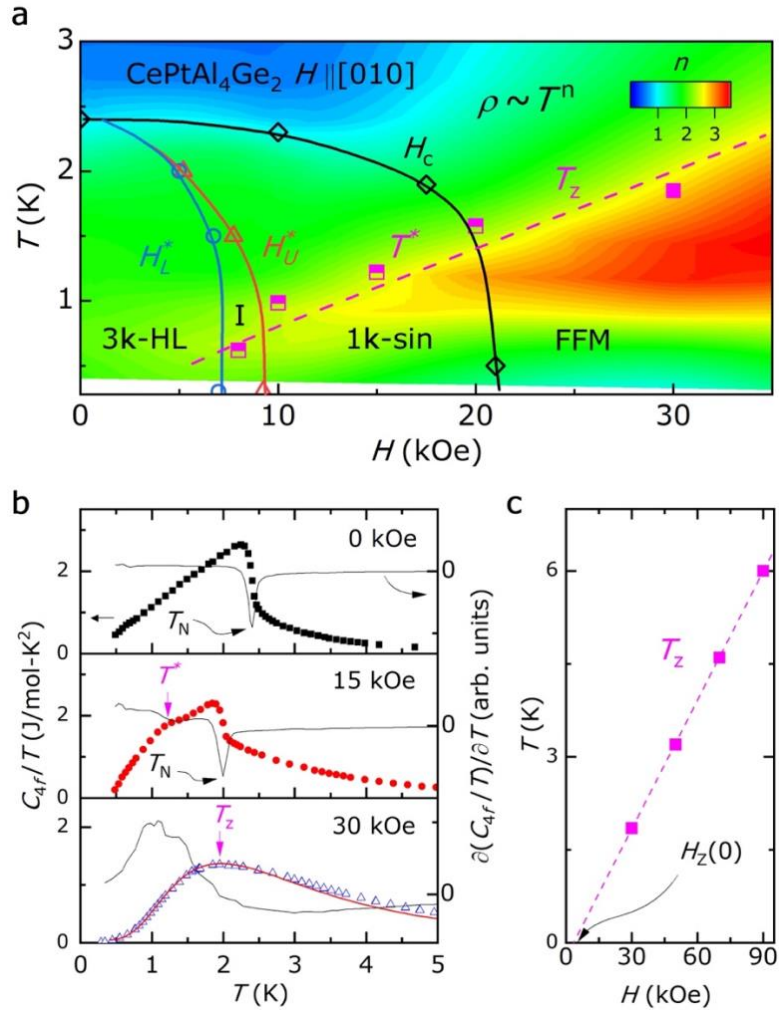

**Fig. 4. Overall $T - H$ phase diagram of CePtAl$_4$Ge$_2$ for $H_{[010]}$ and temperature-dependent magnetic specific heat capacity. a**, Characteristic fields and temperatures are overlaid on the colour contour plot of the temperature exponent of resistivity $n$. **b**, $C_{4f}/T$ and $\partial C_{4f}/\partial T$ are plotted as a function of temperature for three representative fields on the left and right ordinates, respectively, wherein $T_N$ was determined by a dip in $\partial(C_{4f}/T)/\partial T$, while $T_Z$ (closed magenta squares) and $T^*$ (half-filled magenta squares) were assigned as peak temperatures of the specific heat capacity (for other fields, see Fig. S9 in SI). The red line for $H_{[010]} = 30$ kOe case was simulated by the single-ion resonance-level mode of Zeeman splitting for spin-1/2 [47] (for details, see Fig. S10 in SI). **c**, $H_{[010]}$-dependent $T_Z$ shows a linear behaviour at fields higher than $H = 30$ kOe. Here, $H_Z(T=0K) \sim 3.5$ kOe was estimated by linear extrapolation. FFM represents the field-polarized ferromagnetic phase.

# Supplementary Information for
# **Triple-sinusoid hedgehog lattice in a centrosymmetric Kondo metal**

Soohyeon Shin[1,2][†], Jin-Hong Park[3,2][†], Romain Sibille[4], Harim Jang[2], Tae Beom Park[2], Suyoung Kim[5,2], Tian Shang[6,1], Marisa Medarde[1], Eric D. Bauer[7], Oksana Zaharko[4], Michel Kenzelmann[4][*] and Tuson Park[2][*]

[1]Laboratory for Multiscale Materials and Experiments, Paul Scherrer Institut, 5232 Villigen PSI, Switzerland.
[2]Center for Quantum Materials and Superconductivity (CQMS) and Department of Physics, Sungkyunkwan University, Suwon 16419, South Korea.
[3]Research Center for Novel Epitaxial Quantum Architectures, Department of Physics, Seoul National University, Seoul, 08826, South Korea.
[4]Laboratory for Neutron Scattering and Imaging, Paul Scherrer Institut, 5232 Villigen PSI, Switzerland.
[5]Department of Physics, Simon Fraser University, Burnaby, British Columbia, Canada.
[6]Key Laboratory of Polar Materials and Devices (MOE), School of Physics and Electronic Science, East China Normal University, Shanghai 200241, China.
[7]Los Alamos National Laboratory, Los Alamos, NM 87545, USA.

*Corresponding author(s). E-mail(s): michel.kenzelmann@psi.ch; tp8701@skku.edu;
[†]These authors contributed equally to this work.

# 1 Detailed analysis based on multi-k spin model

## 1.1 Multi-k states

Here, we provide the details of spin structure analysis based on multi-**k** spin model. Before directly going into the theoretical description, let us first summarize the experimental situation in the main text. As external magnetic fields along $H_{\|[010]}$ increase, the phase transitions occur as evidenced by Fig. 1c. More precisely, the neutron diffraction measurement in Fig. 3b indicates that it is related to the phase transition from the multi-arm of spin state to single-arm of spin state. The field dependence of **k**-peak from neutron scattering measurement provides the evidence of the well-defined multi-**k** spin structure as pointed out in the main text. In this multi-arm of spin description, the three arms of spin states are degenerate at zero magnetic fields, motivating us to use the superposition of multi-**k** spin model. Previously, the multi-**k** states are studied in the various spin states. The well-known example of the multi-**k** states is the skyrmion lattice states in the chiral magnet [1], realizing the superposition of spiral spin states with three different propagating directions. The skrymion is topologically distinct object as illustrated in Fig. S1a. In the context of chiral magnet, **k** represents the spin propagating vector, meaning that the direction of spin is slightly shifted along the **k** vector with the amplitude of spin intact. Although a few examples of $f$-electron materials [2] and $J_1 - J_2$ frustrated magnets [3] also exhibit the multi-**k** states, all the **k**'s represent the spin propagating vectors. In this study, we want to stress that the *spin amplitude modulation* vectors **k** can also be superposed in form of multi-**k** states. With this in mind, we construct the spin structures with multi-**k** states. The spin configurations are given by

$$\mathbf{S}(\mathbf{r}) = \sum_{\alpha=1}^{3} \sqrt{1 - \cos\left(\frac{2\pi}{p}\mathbf{k}_\alpha \cdot \mathbf{r}\right)} \left(e^{i\pi\mathbf{k}_\alpha \cdot \mathbf{r}}(i\mathbf{k}_\alpha) + c.c\right). \quad (1)$$

Here $p$ denotes the periodicity for a given lattice size $L$, so that one can freely tune the periodicity to mimic the spin structure from experimental data. The overall factor $\sqrt{1 - \cos\left(\frac{2\pi}{p}\mathbf{k}_\alpha \cdot \mathbf{r}\right)}$ encodes the amplitude modulation where $\mathbf{k}_\alpha$ ($\alpha = 1, 2, 3$) is the unit vectors. The term of $e^{i\pi\mathbf{k}_\alpha \cdot \mathbf{r}}$ indicates the antiferromagnetic order along $\mathbf{k}_\alpha$. The spins are pointed to the directions of the amplitude modulation vector, which is captured by the expression of $i\mathbf{k}_\alpha$ in Eq. (1). With this spin formula, we have successfully reproduced the spin configurations of Fig. S2.

## 1.2 Energetic stability of spin structure

In this section we discuss the energetic stability of spin structures, which are constructed by using spin formula of Eq. (1) and individual single-$\mathbf{k}_\alpha$ modes ($\alpha = 1, 2, 3$). The minimal Hamiltonian to calculate the energy of spin structure is $H = H_J + H_Z$, where $H_J$ is Heisenberg-type interactions $H_J = J \sum_{ij} \mathbf{S}_i \cdot \mathbf{S}_j$ with the leading order of $J$ while $H_Z$ is Zeeman coupling interaction. The energy calculation is done on the triangular lattice. Since the spins are coupled with magnetic fields through Zeeman coupling, the energetic advantage is taken by the intrinsic spin structures whether the spins are aligned with magnetic fields or not. Figure S3 shows the resultant energy calculation. At zero magnetic field, the energy of single-$\mathbf{k}_\alpha$ is almost degenerate showing $E_{\mathbf{k}_1} = -0.1105J$, $E_{\mathbf{k}_2} = -0.1067J$, $E_{\mathbf{k}_3} = -0.1106J$. The energy degeneracy is slightly shifted since the energies are calculated in the finite system. We can also calculate the energy of multi-$\mathbf{k}$ spin structure as $E_{\text{multi}-\mathbf{k}} = -0.1345J$. The energy difference is $\Delta E = E_{\text{multi}-\mathbf{k}} - E_{\text{single}-\mathbf{k}} \sim -0.02J$. This slight energetic advantage of multi-$\mathbf{k}$ over single-$\mathbf{k}_\alpha$ mode can be explained in a way that the multi-$\mathbf{k}$ state respects geometrical frustration more than single-$\mathbf{k}_\alpha$ state. As we increase the magnetic field along [010]-direction, we observe the phase transition from multi-$\mathbf{k}$ state to single-$\mathbf{k}_\alpha$ spin phase as indicated in Fig. S3.

### 1.3 Topological property of multi-k states

Here we discuss the topological aspect of spin structures. Figure S4a demonstrates the magnetic structures simulated by Eq. (1), and the spin hedgehog around the singular point is shown in Fig. S4b. The periodic vanishing of spin moments leads to us to consider a topological invariance, namely hedgehog number. This can have nonzero value around a singular point. For comparison with skyrmion structure, skyrmion has no singular point at the center while hedgehog spin state has singular point at the center of spin structure. In general, there are several ways to measure the hedgehog number. For the sake of convenience we would obtain hedgehog number from the calculations of the skyrmion (winding) number for a fixed plane. In the followings we provide the detailed calculations. After the coordinate transformation $x' = \frac{\sqrt{3}}{2}y - \frac{81}{1250}z$, $y' = \frac{\sqrt{3}}{2}x + \frac{81}{1250}z$, and $z' = -\frac{\sqrt{3}}{2}x + \frac{\sqrt{3}}{2}y + \frac{81}{1250}z$, we can re-write Eq. (1) as

$$\mathbf{S}(\mathbf{r}') = \sum_{\alpha=1}^{3} \mathbf{S}_{\mathbf{k}_\alpha}(\mathbf{r}'), \qquad (2)$$

where

$$\begin{aligned}
\mathbf{S}_{\mathbf{k}_1}(\mathbf{r}') &= \sqrt{1 - \cos\left(\frac{2\pi}{p}y'\right)} \sin \pi y' \left(\frac{\sqrt{3}}{4}, \frac{1}{4}, \frac{81}{2500}\right), \\
\mathbf{S}_{\mathbf{k}_2}(\mathbf{r}') &= \sqrt{1 - \cos\left(\frac{2\pi}{p}x'\right)} \sin \pi x' \left(0, -\frac{1}{2}, \frac{81}{2500}\right), \\
\mathbf{S}_{\mathbf{k}_3}(\mathbf{r}') &= \sqrt{1 - \cos\left(\frac{2\pi}{p}z'\right)} \sin \pi z' \left(-\frac{\sqrt{3}}{4}, \frac{1}{4}, \frac{81}{2500}\right).
\end{aligned} \qquad (3)$$

As indicated in the above expressions, the spin texture is also slowly varied along the $z$-axis as in Fig. S4c. After restoring $\mathbf{r}' \to \mathbf{r}$ we expand the spin $\mathbf{S}(\mathbf{r})$ around the particular singular point $\mathbf{r}_0$ giving the magnetization as follows:

$$\mathbf{S}(\mathbf{r} + \mathbf{r}_0) \approx \mathbf{S}(\mathbf{r}_0) + \mathbf{S}'(\mathbf{r}_0)\mathbf{r} + O(\mathbf{r}^2) \equiv \boldsymbol{m}(r). \tag{4}$$

The linearized expressions are given by

$$\begin{aligned} m_x(\mathbf{r}) &= \frac{\sqrt{3}}{2\sqrt{2}p}(a_y y - a_z z)/|\boldsymbol{m}(r)|, \\ m_y(\mathbf{r}) &= \frac{1}{2\sqrt{2}p}(-2a_x x + a_y y + a_z z)/|\boldsymbol{m}(r)|, \\ m_z(\mathbf{r}) &= \frac{81}{1250\sqrt{2}p}(a_x x + a_y y + a_z z + m_0)/|\boldsymbol{m}(r)|, \end{aligned} \tag{5}$$

with $a_i = \frac{\sqrt{2}\pi}{p}\left(\cos\left(\frac{\pi i_0}{p}\right)\sin(\pi i_0) + p\cos(\pi i_0)\sin\left(\frac{\pi i_0}{p}\right)\right)$ where $i = (x, y, z)$. Here the uniform magnetization $m_0$ is included in the $z$-component, which is due to the magnetic field. The resulting skyrmion number defined by $N_{\text{sk}} = 1/4\pi \int d^2\mathbf{r} \frac{1}{\mathbf{m}^3}\mathbf{m}(\mathbf{r}) \cdot (\partial_x\mathbf{m}(\mathbf{r}) \times \partial_y\mathbf{m}(\mathbf{r}))$ (41) are calculated as follows:

skyrmion density =

$$-\frac{1}{4\pi} \frac{(m_0 + c_0 z)}{\left(c_1(a_y y - a_z z)^2 + c_2(-2a_x x + a_y y + a_z z)^2 + c_3(m_0 + a_x x + a_y y + a_z z)^2\right)^{3/2}}, \tag{6}$$

where $c_i$ is some constants. Figure S4d shows that the skyrmion number changes from $+0.5$ to $-0.5$ when $m_0$ passes from positive to negative. By the definition of hedgehog number $N_{\text{h}} = N_{\text{sk}}(-m_0) - N_{\text{sk}}(+m_0)$, $N_{\text{h}}$ at the singular point is $-1$ [4]. This can support the idea that the topological phase transition is associated with the evolution of hedgehog number from the multi-$\mathbf{k}$ spin state to single-$\mathbf{k}$ spin state as in-plane magnetic fields increase.

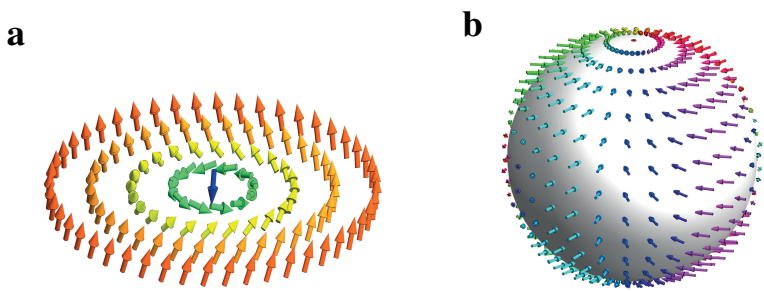

**Fig. S1 Spin configurations of the skyrmion and hedgehog are illustrated. a**, The skyrmion spin structure is that three spin components are projected in two-dimensional space. **b**, The hedgehog spin structure is that three spin components are projected in three-dimensional space with singular point at center.

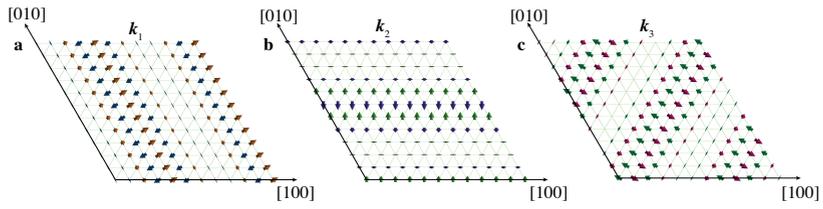

**Fig. S2 Simulated spin configurations.** Spin configurations of Eq. (1) with the propagating vectors of $\mathbf{k}_1$, $\mathbf{k}_2$, and $\mathbf{k}_3$ on the triangular lattice basis are depicted in **a-c**.

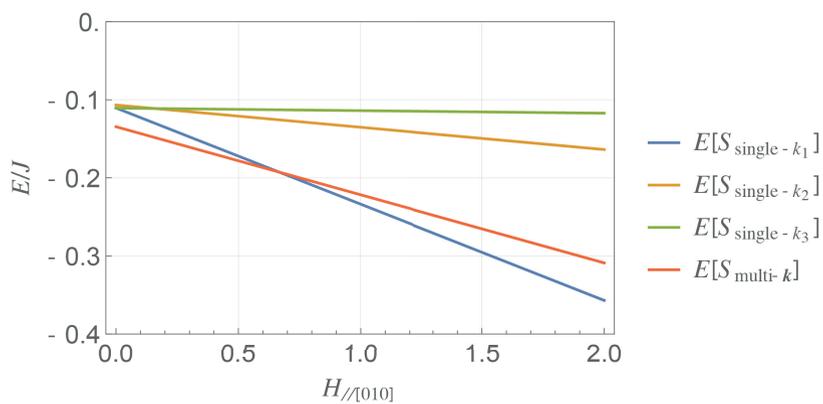

**Fig. S3 Energy calculations in unit of Heisenberg interaction of $J$.** The calculation shows that the energetic preference is changed from multi-$\mathbf{k}$ to single-$\mathbf{k}_1$ state as the magnetic field along [010] increases.

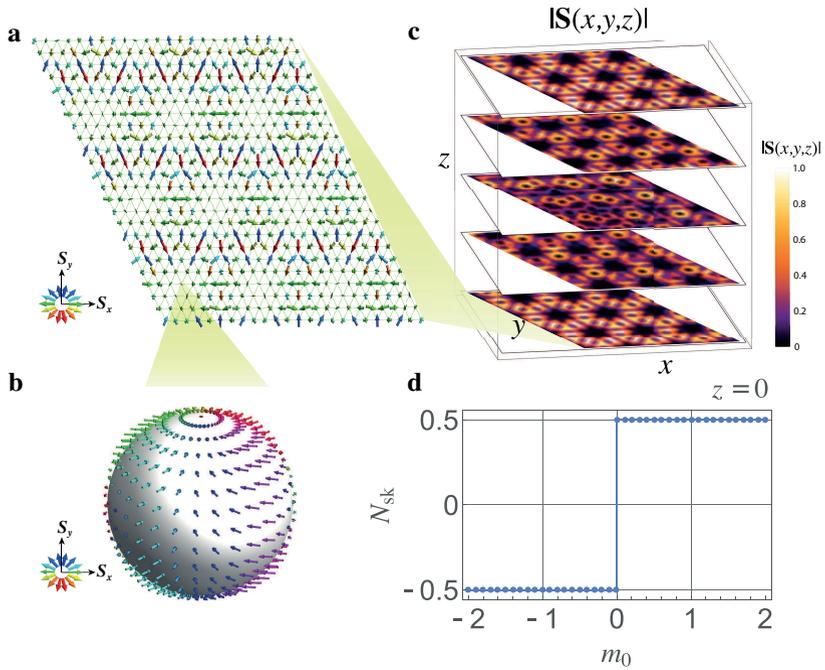

**Fig. S4 Simulation of multi-k spin structure and calculation of topological hedgehog number.** Spin textures for multi-**k** structure are illustrated in **a**. The hedgehog spin texture around singular points is depicted in **b**. The evolution of the spin texture along $z$ is encoded by $|\mathbf{S}(x,y,z)|$ in **c**, which shows the slowly varying spin profile along the $z$-axis as well as within the $xy$-plane, justifying the existence of the hedgehog spin structure. The change in the skyrmion number as a function of the uniform magnetisation $m_0$ on the $xy$-plane at $z=0$ is shown in **d**.

## 2 Supplementary experimental results

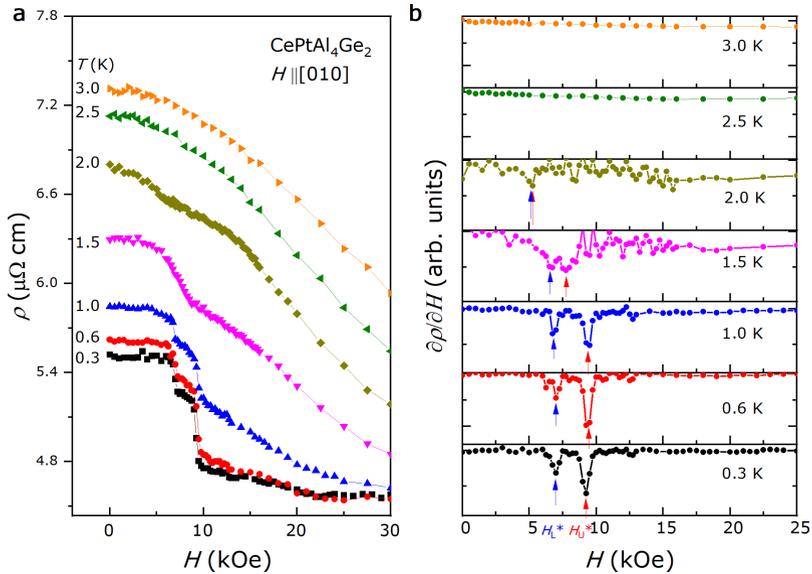

**Fig. S5 Field-dependent electrical resistivity ($\rho(H)$) and first derivative of resistivity ($\partial\rho/\partial H$) of CePtAl$_4$Ge$_2$ a**, $\rho(H)$ data at $T = 0.3$ K exhibits two drops with increasing magnetic fields applied along the crystallographic [010] direction. The two drops become weaker and shift to the lower fields with increasing temperature. **b**, Field-dependent $\partial\rho/\partial H$ shows two distinct peaks corresponding to the resistivity drops in $\rho(H)$, wherein two characteristic fields ($H_L^*$ and $H_U^*$) are indicated as shown in figure.

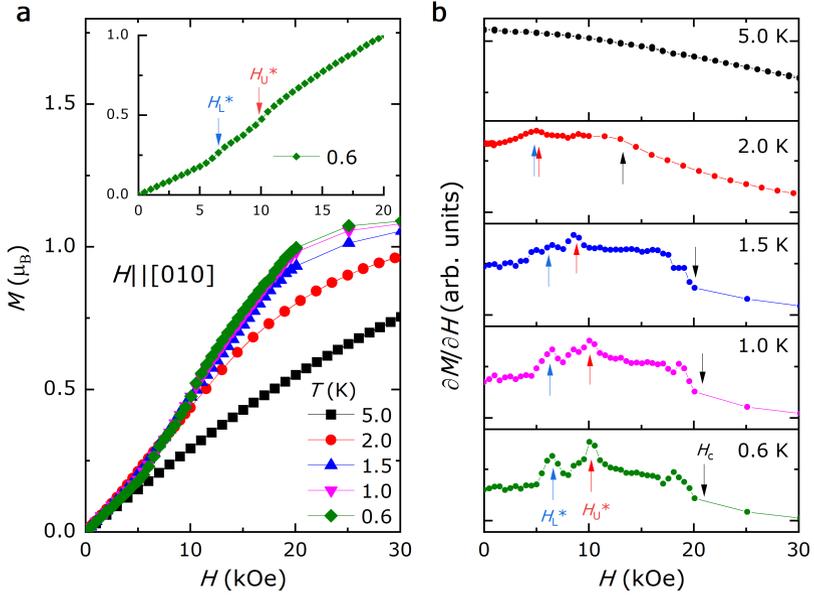

**Fig. S6 Field-dependent magnetisation ($M(H)$) and first derivative of magnetization ($\partial M/\partial H$) of CePtAl$_4$Ge$_2$. a**, $M(H)$ data were measured at various temperatures under magnetic fields applied along the crystallographic [010] direction. Inset to **a**, shows the data at $T = 0.6$ K in detail. The two small jumps of $M(H)$ at $T = 0.6$ K correspond to the two resistivity drops at $H_L^*$ and $H_U^*$ **b**, Field-dependent $\partial M/\partial H$ exhibits two distinct peaks corresponding to two small jumps in $M(H)$. $H_c$ indicates the antiferromagnetic phase boundary.

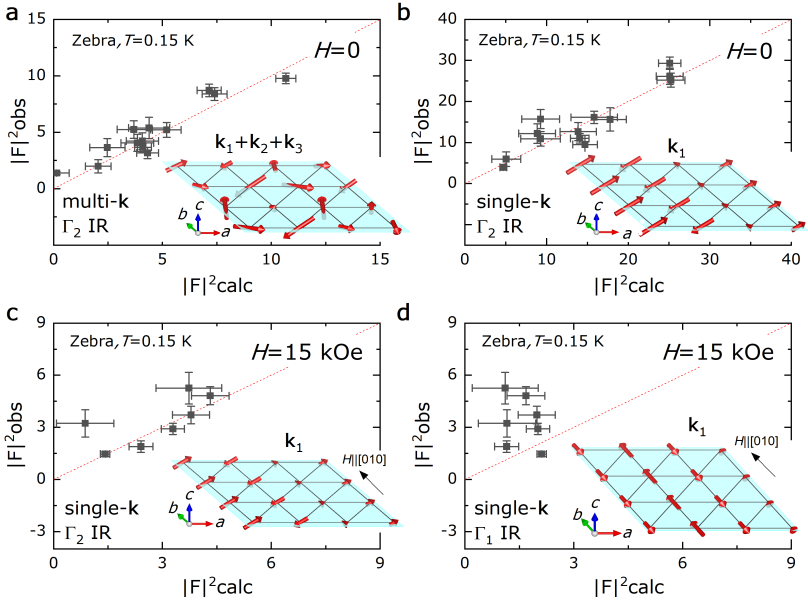

**Fig. S7 Single-crystal neutron diffraction data and refinement results of CePtAl$_4$Ge$_2$.** Zero-field magnetic reflections were analyzed using a multi-**k** and multi-domain structure with $\Gamma_2$ irreducible representation (IR) as shown in **a** and **b**, respectively. Magnetic reflections measured at 15 kOe were analyzed using a single-**k** structure with $\Gamma_2$ and $\Gamma_1$ IR as shown in **c** and **d**, respectively. Inset to each panel shows the deduced magnetic structure from the refinement. Note that $|F|^2$obs(calc) indicates the observed (calculated) integrated neutron intensity, when the F is the magnetic structure factor.

**Table S1** Refinement results on the magnetic structures of CePtAl$_4$Ge$_2$ at $H_{[010]} = 0$ and 15 kOe.

| $H_{[010]}$ | Model | $\mathbf{m}_{\text{in-plane}}$[1] | $\mathbf{m}_{[001]}$ | $R_F$ |
|---|---|---|---|---|
| 0 | multi-**k** $\Gamma_2$ | 1.49(3) | 0.35(4) | 7.38 |
| 0 | multi-domain $\Gamma_2$ | 1.29(3) | 0.21(4) | 8.72 |
| 15 kOe | single-**k** $\Gamma_2$ | 0.65(3) | 0.13(2) | 13.30 |
| 15 kOe | single-**k** $\Gamma_1$ | 0.36(4) | - | 33.70 |

[1]$\mathbf{m}_{\text{in-plane}}$ is $[210]/\sqrt{3}$ for $\Gamma_2$ and [010] for $\Gamma_1$ IR. The unit of **m** is $\mu_B$.

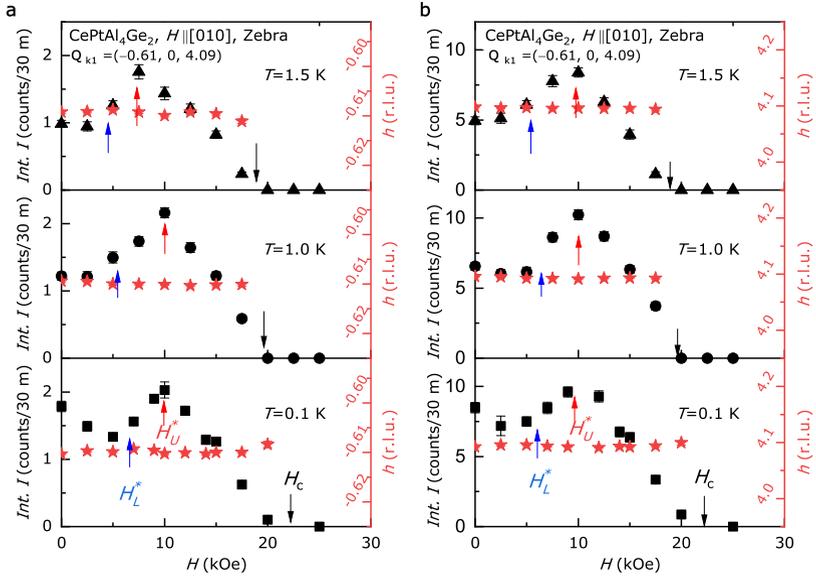

**Fig. S8 Field-dependent integrated neutron intensity of CePtAl$_4$Ge$_2$.** $H_{[010]}$-dependent integrated neutron intensity of magnetic reflection (-0.61, 0, 4.09) at $T = 0.1$, 1.0, and 1.5 K for $h$- (**a**) and $l$- (**b**) scans. Integrated neutron intensity was estimated using Gaussian fit to each scan. Characteristic fields are indicated by blue, red, and black arrows for $H_L^*$, $H_U^*$, and $H_c$, respectively. Error bars are smaller than symbols.

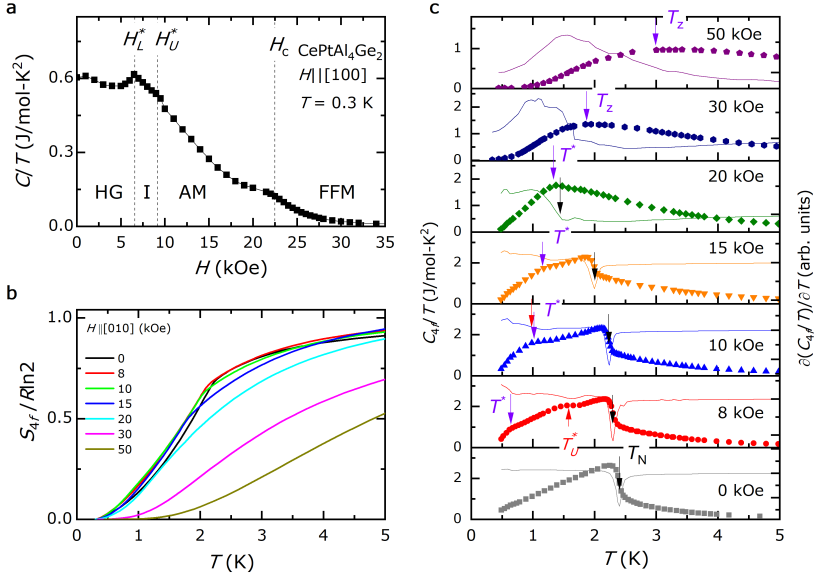

**Fig. S9 Field- and temperature-dependent $C/T$, $C_{4f}/T$, $\partial(C_{4f}/T)/\partial T$, and $S_{4f}$ of CePtAl$_4$Ge$_2$. a**, $H_{[010]}$-dependent $C/T$ was measured at $T = 0.3$ K, and three black dashed lines separate the four magnetic regimes (hedgehog (HG), intermediate (I), amplitude modulation (AM), and field-polarized ferromagnet (FFM) states). **b**, Magnetic entropy divided by $R\ln 2$ is plotted as a function of temperature for various $H_{[010]}$. **c**, $C_{4f}/T$ and $\partial(C_{4f}/T)/\partial T$ measured at various $H_{[010]}$ are plotted as a function of temperature on the left and right ordinates, respectively. The broad humps in $C_{4f}/T$ correspond to $T_Z$, $T^*$, and $T_U^*$. $T_N$ is shown as a sharp dip of $\partial(C_{4f}/T)/\partial T$. $T_U^*$ corresponds to the $H_U^*$.

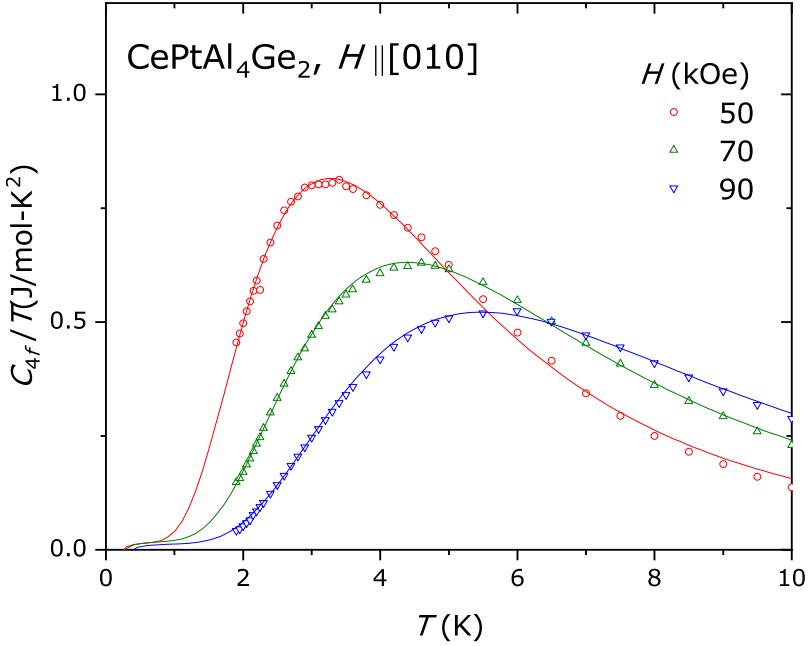

**Fig. S10 Temperature-dependent $C_{4f}/T$ for various $H_{[010]}$ larger than the critical field of AFM phase.** Each solid line represents the calculation based on the single-ion resonance-level mode Zeeman splitting for spin-1/2, which is $C_{4f}/T = k_B\Delta/(\pi k_B T^2) - 2(k_B/T)Re[(\Delta+iE)^2/(2\pi k_B T)^2(4\Psi'(1+(\Delta+iE)/(\pi k_B T))) - \Psi'(1+(\Delta+iE)/(2\pi k_B T))]$ (37). Here, $k_B$ is the Boltzmann constant, $\Delta = k_B T_K$ ($T_K$ is the Kondo temperature) related with the broadening of the Schottky anomaly of $C_{4f}/T$, $E = g_J\mu\mu_0 H$ ($g_J$ is the Lande g-factor, $\mu$ is the size of the magnetic moment, $\mu_0$ is the magnetic permeability), which is the Zeeman splitting energy of the doublet ground state, and $\Psi'$ is the derivative of the digamma function. Table S2 summarizes the fitting results for $H_{[010]} = 50$, 70, and 90 kOe, where the fields are high enough for single-ion resonance-level mode.

**Table S2** $T_K$, $E$, and $T_{max}$ parameters deduced from the fittings using single-ion resonance-level mode Zeeman splitting for $H_{[010]} = 50$, 70, and 90 kOe. $T_{max}$ represents the temperature where $C_{4f}/T$ shows maximum or $\partial(C_{4f}/T)/\partial T = 0$.

| $H_{[010]}$ (kOe) | $T_K$ (K) | $E$ (K) | $T_{max}$ (K) |
|---|---|---|---|
| 50 | 0.20 | 10.67 | 3.2(1) |
| 70 | 0.41 | 14.41 | 4.6(1) |
| 90 | 0.44 | 17.91 | 6.0(1) |

# References

[1] Yu, X.Z., Onose, Y., Kanazawa, N., Park, J.-H., Han, J.H., Matsui, Y., Nagaosa, N., Tokura, Y.: Real-space observation of a two-dimensional skyrmion crystal. Nature **465**, 901–904 (2010) https://doi.org/10.1038/nature09124

[2] Marcus, G.G., Kim, D.-J., Tutmaher, J.A., Rodriguez-Rivera, J.A., Birk, J.O., Niedermeyer, C., Lee, H., Fisk, Z., Broholm, C.L.: Multi-$q$ Mesoscale Magnetism in CeAuSb$_2$. Physical Review Letters **120**, 097201 (2018). https://doi.org/10.1103/PhysRevLett.120.097201

[3] Leonov, A.O., Mostovoy, M.: Multiply periodic states and isolated skyrmions in an anisotropic frustrated magnet. Nature Communications **6**, 1–8 (2015). https://doi.org/10.1038/ncomms9275

[4] Park, J.-H., Han, J.H.: Zero-temperature phases for chiral magnets in three dimensions. Phys. Rev. B **83**, 184406 (2011). https://doi.org/10.1103/PhysRevB.83.184406